\documentclass[prl,showpacs,twocolumn,aps,superscriptaddress]{revtex4}

\usepackage{graphicx}
\usepackage{bm}
\usepackage{amssymb}
\usepackage{amsmath}
\usepackage{subfigure}
\usepackage{tikz}

\begin{document}

\title{Unveiling the physics of the doped phase of the Kagome $t-J$ model}
\author{Siegfried Guertler}
\affiliation{Bethe Center for theoretical physics, The University of Bonn, Bonn, Germany}
\affiliation{Lehrstuhl f\"ur Theoretische Physik II, Technische Universit\"at Dortmund, 44221 Dortmund, Germany}
\author{Hartmut Monien}
\affiliation{Bethe Center for theoretical physics, The University of Bonn, Bonn, Germany}
\date{\today }

\begin{abstract}
We investigate the ground state properties of the Kagome lattice $t-J$ model at low doping. We propose a new state of matter. Our results suggest an interesting phase with a balance of spin-exchange and kinetic
exchange through the building blocks of stars and their internal hexagons, being linked by triangles. The particles favor to hop in the hexagons, while the spin-exchange is taking place mainly on the stars. The combined effect leads to the formation of the ``Star of David" valence bond crystal. We discuss our result in connection with static impurities, and show the likely relevance to the diluted Kagome lattice Heisenberg model, describing actual compounds.
\end{abstract}

\maketitle

{\it Introduction: } The ground state of the Kagome Heisenberg model which has been studied for several decades is still an unresolved problem. Due to the high degeneracy and 
competitive ground state energies, different methods lead to the proposal of very different ground states. It is not clear yet, how the choice of  
boundary conditions and symmetry, usually not a severe problem for other models, would bias the result. Pushing numerical accuracy and performing larger and larger scale computations 
has been the main route to address this question recently \cite{white,iqb,HUS1,LAEU,DEP}. Unfortunatly the various possible methods did not reach a consistent conclusion yet.
For a real material the situation would differ, as there would be many sources of perturbations lifting degeneracies.
On the other hand experimentally it is clear, that compounds with a Kagome lattice do possess very interesting properties, even though there are these perturbations, suggesting that the interesting physics
survives in the perturbed case. Such possible perturbations are Dzyaloshinski-Moriya interaction, long-range interactions, anisotropy and impurities \cite{RIG,NAR,HIROI,OLA}. Specifically dynamic impurities are an interesting question which have not been investigated in detail. E.g. the so far best candidate of an ideal Kagome lattice, being ZnCu$_3$(OH)$_6$Cl$_2$, has very interesting experimental properties while impurities are present \cite{IMAI,Mendels1,OLA,shores1,VRIES,Helton1}.\\
The study of doping is interesting due to several aspects: (i) There is a connection between spin-liquids and superconductivity, as can be seen in the square lattice $t-J$ model where
the doped phase is a RVB-spin liquid possessing off-diagonal long-range order. (ii) Real materials such as Zn-paracatamite have impurities. In Zn-paracatamite these impurities are most likely static, yet we choose the dynamic version, as we believe the physics involved is more interesting. We discuss and argue our works impact on the static case below, and suggest that the two models are not completely independent. (iii) As discussed in a previous reference \cite{GUERT}, the doped phase appears to favors some sort of VBC. For static impurities the trend to dimerization has been shown in 
\cite{DOM}. It was surprising at first to see this tendencies in the dynamic version, in this paper we point out the likely reasons.\\
In an earlier study by us, we investigated low doping systematically up to a doping level of about 30 percent, testing several wave-function in that range. In this investigation we study the doping level of about 10 percent. We compared the best wave-function of our earlier study with alternative versions of VBCs and refined the original VBCs with a further parameter to improve it. The mean-field 
unit cell is in most cases relatively large, therefore making it hard to study correlation functions as there are many in-equivalent sites. There is a computational limit on the system-size, systematic correlation-functions over distance are therefore currently difficult to obtain. In this study we focus on local bond observables, revealing some of the physics involved.\\

{\it Model and Method: } We study the $t-J$ model within the subspace of single occupied lattice sites:

\begin{eqnarray}
H&=&-t\sum_{\langle ij\rangle}\sum_{\sigma=\uparrow\downarrow}\mathcal{P}\left(c^{\dagger}_{i\sigma}c_{j^{\phantom\dagger} \sigma}+ h.c.\right)\mathcal{P}+ \nonumber \\
 & & +J\sum_{\langle ij\rangle}\left(\mathbf{S}_i\cdot\mathbf{S}_j-\frac{1}{4}n_i n_j\right)
\label{ham}
\end{eqnarray}

\begin{figure}[h]
   \begin{center}
    \begin{tabular}{cc}
    \resizebox{37mm}{!}{\includegraphics[width=0.80\columnwidth]{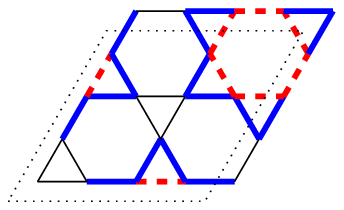}} &
    \resizebox{37mm}{!}{\includegraphics[width=0.80\columnwidth]{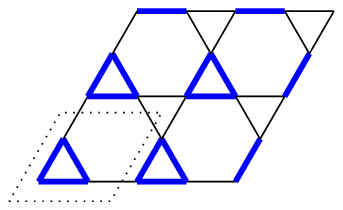}}\\
    \resizebox{37mm}{!}{\includegraphics[width=0.80\columnwidth]{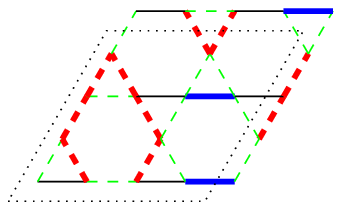}}
    \end{tabular}
   \end{center}
\caption{(Color online) The parametrisation of variational VBC-states considered. All fluxes are zero. Above left: D-VBC where the blue thick bonds correspond to the VBC of Hastings and are controlled by $\chi_1$, the red dashed bonds is the parametrization of the inner hexagon and their strength is controlled by $\chi_2$. The remaining bonds we refer to as 'weak' bonds. Above right: I-VBC as suggested by Indergand et al for the 1/3 filling case, where the blue thick bonds are controlled by a variational parameter $\chi_1$ and the weak bonds (appearing in black thin lines) are set to unity. Below: Y-VBC which was found as perturbation of the Kagome lattice Heisenberg model in DMRG. The thick blue bonds are the strongest bonds controlled by a variational parameter $\chi_1$, the ``8-site loop" is indicated by thick dashed red lines, and is controlled by a parameter $\chi_2$, the weakest bonds are depicted as thin dashed green lines controlled by a parameter $\chi_3$, and the remaining connecting bonds in thin solid black lines, are set to unity.}
\label{VBCS}
\end{figure}

Here $c_{j\sigma}$ is the electron annihilation operator of an
electron with spin $\sigma$ on site $i$, $\vec S_i$ is the spin-1/2 operator at site $i$, and
$n_{i\sigma}=c_{i\sigma}^{\dagger} c_{i\sigma}$. The sum $\langle i,j \rangle$
is over the nearest neighbors (n.n.) pairs on the Kagome lattice. $\mathcal{P}=\prod_{i}(1-n_{i\uparrow}n_{i\downarrow})$ is the projection operator enforcing the single occupancy constraint. We fix $J=0.4t$ in what follows.\\ 
Our starting point is the wave-function of the uniform state (U-state) with no flux through the triangles and hexagons, in combination with the valence bond crystal state suggested by Hastings (H-VBC-state), with a 12-site unit-cell \cite{HAS1}. This was the best wave-function found in an earlier study by us \cite{GUERT}. Motivated partly by an intermediate result of local measurements, here we use an improved wave-function  
using an additional variational parameter, which we call the D-VBC state (``Star of David"). The underlying flux is zero for all plaquettes, the VBC patten consists of three types of bonds: strong bonds, forming the outer shape of the ``Star of David", weak bonds, in the inner hexagon within the ``Star of David", and connecting bonds, which we refer to all other bonds being neither strong nor weak. The value $\chi_1$ in bracket parametrizes the strength of the bond-modulation for the strong bonds, as the amplitude ratio between the Star of David VBC bonds and the connecting bonds. Independent of $\chi_1$ we improve this wave-function by varying the bond strength of the weak bonds as well of the inner hexagon loop, this parameter is called $\chi_2$. This quantity parametrizes the amplitude ratio between the inner hexagon bonds of the Star of David and the connecting bonds. We tried two alternative patterns of VBCs found by other investigations: Indergand et all \cite{INDE} suggested a state where all the up and down triangles decouple at the fillings of 1/3 and 2/3 while the authors of the paper argue it to be a state slightly distinct from a VBC being dubbed ``bond-order-wave" we believe that our wave-function should be able to capture this state by fixing
amplitude ratios on the bonds in question. This state has been proposed for another doping level than the one considered here, nevertheless we tested it, as normally one would expect a stable phase to be realized over a wider range.  In addition the state has been argued as a result of a ``one hole in three sites" condition on the elementary triangles, it therefore appears reasonable that at doping levels related to this ratio, such as 1/6 or 1/12 similar physics may appear. We call this state I-VBC here. On the other hand at half filling the recent DMRG investigation by Yan et al \cite{white} found a particular patten which could be argued to be closely related to the spin-liquid at this point. Suggesting the spin-liquid at half-filled of being a melted version of this type of VBC. To see if the introduction of doping favors this state we tested it at our doping level. We refer to this state as Y-VBC.
See Fig. \ref{VBCS} for a visualization of our parametrization.
We use a standard VMC scheme and scan the parameter-space by the mesh-method. This has advantages for parallelization and for exploratory studies one can systematically study single parameter contributions, which are lost in an automatic search. In the case of the Kagome lattice it is hard to find a scheme of boundary conditions which avoids all degenerate states for varying lattice sizes and fillings. We therefore chose to work with periodic boundary conditions (PBC) in all cases. In addition antiperiotic boundary condition might be problematic when studying valence bond crystals. Typically we performed around 20.000 to 100.000 sweeps to equilibriate the system and another 50.000 to 400.000 sweeps for measurements. We then average our data over 8-64 independent runs, to estimate the error. In our investigation we used initially lattices from $N=192$ to $N=432$ sites. We found that $N=192$ sites appears already sufficient.\\

\begin{figure}[h]
     \centering
\includegraphics[width=0.50\columnwidth]{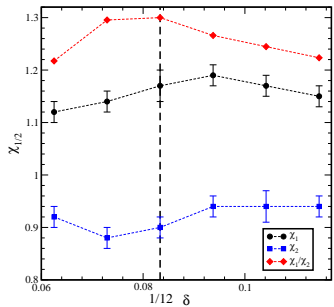}
\caption{Variational parameters for the D-VBC, and the ratio between the two. Both parameters develop their maximum/minimum very close to our suggested optimal doping ($\d=1/12$) for this state (indicated by a dashed vertical line). The ratio between the two has the maximum exactly at this point.}
\label{opt}
\end{figure}

\begin{figure}[h]
   \begin{center}
    \begin{tabular}{cc}
    \resizebox{43mm}{!}{\includegraphics[width=1.00\columnwidth]{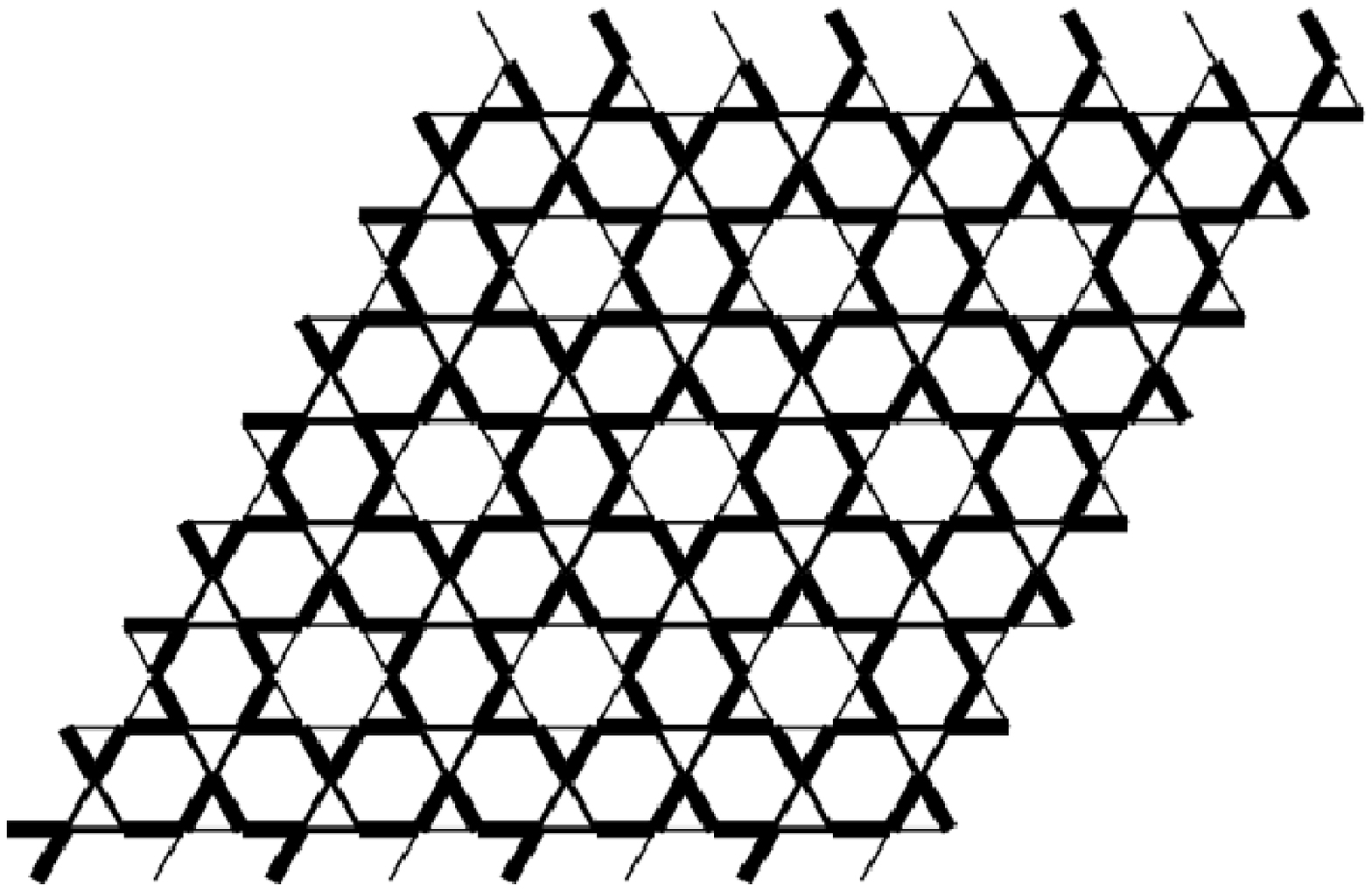}} &
    \resizebox{43mm}{!}{\includegraphics[width=1.00\columnwidth]{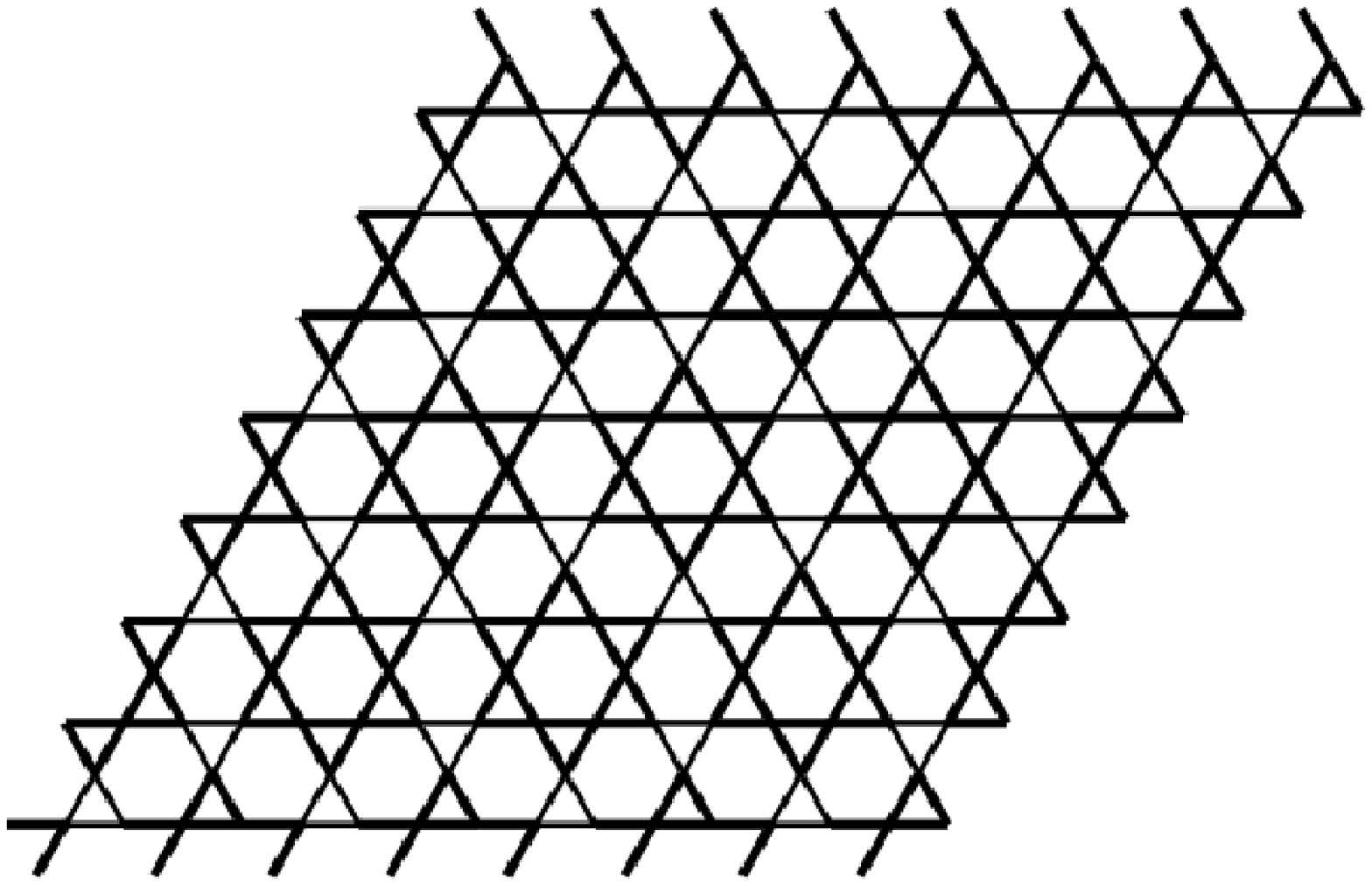}}\\
    \resizebox{43mm}{!}{\includegraphics[width=1.00\columnwidth]{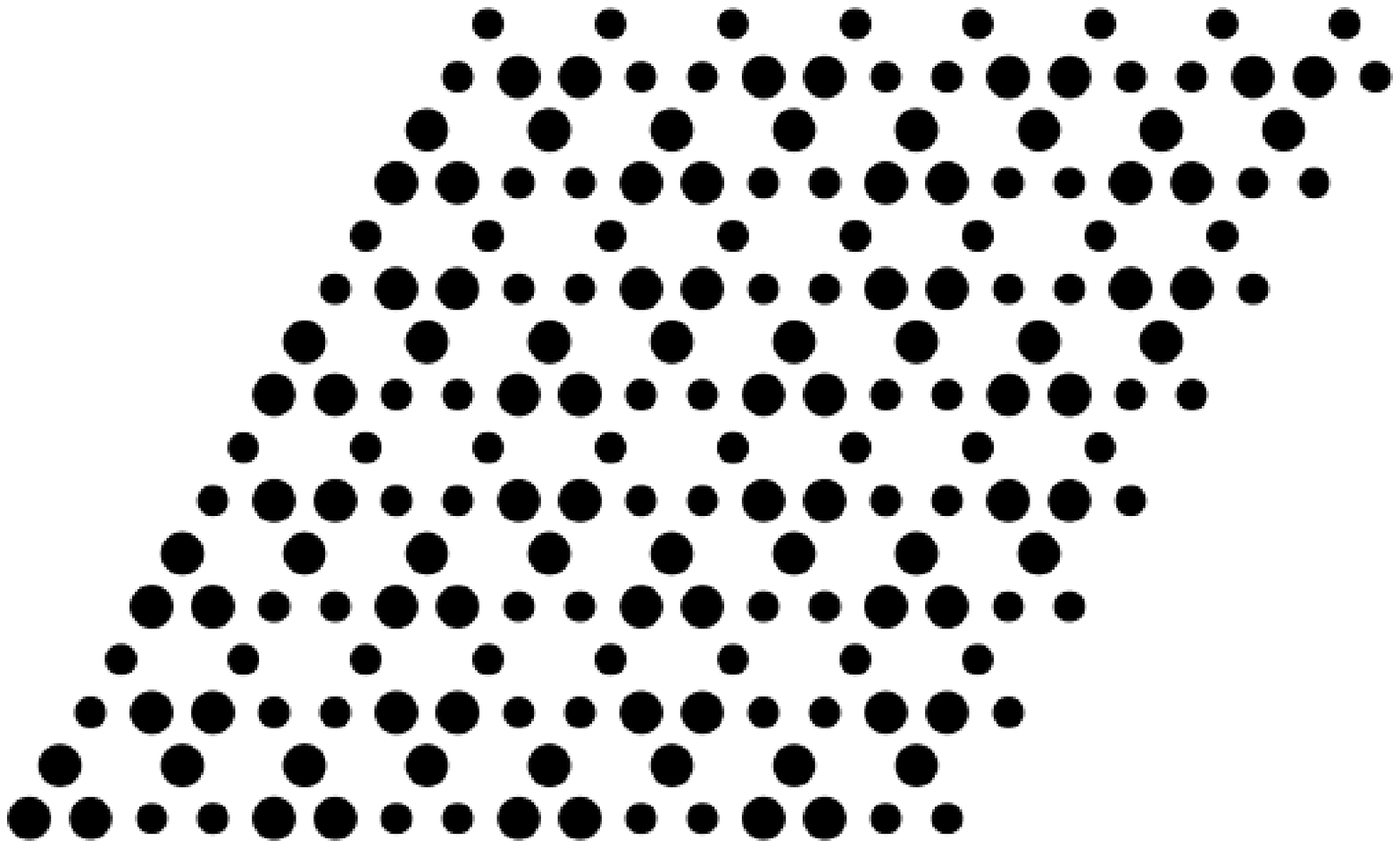}}
    \end{tabular}
   \end{center}
\caption{Local bond and site measurements in the optimized state for a $N=192$ sites system and a doping level of 9.4 percent. Above left: Bond spin exchange; strong bonds have a value of $S_{ij}\approx -0.30$, medium ones are in the range $-0.14<S_{ij}<-0.11$ and the weakest bonds have $S_{ij}\approx -0.05$. Above right: Bond kinetic energy; strong bonds have a value of in the range $-0.14<T_{ij}<-0.15$ weak bonds have $T_{ij}\approx -0.09$. Below: Site densities; For sites with a large density we measure $n_i\approx 0.92$ while the lower density sites have $n_{i}\approx 0.90$.
A finite value have been substracted from all bonds/sites to visualize the pattern better.}
\label{SPINKINDENS}
\end{figure}

\begin{figure}[h]
     \centering
    \begin{tabular}{c}
   \resizebox{41mm}{!}{\includegraphics[width=0.80\columnwidth]{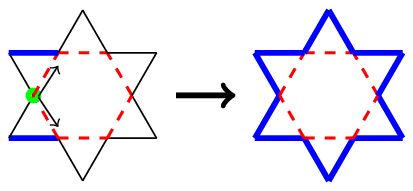}}  \\
   \resizebox{34mm}{!}{\includegraphics[width=0.80\columnwidth]{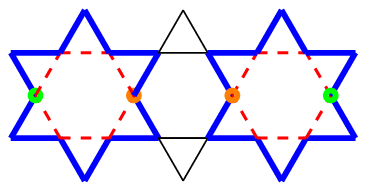}}
    \end{tabular}
\caption{Above: Mechanism forming the ``Star of David": When a hole moves within the inner hexagon completing a route, the outer ``Star of David" is formed. Below: Distribution of Holes: Assuming one hole per star, we show the position of closest and largest distance.}
\label{DIM}
\end{figure}

{\it Results: } The two alternative patterns for VBC states the Y-VBC and the I-VBC have slightly better energies compared to the uniform (zero-flux) state, but these two pattens have higher energies compared to the H-VBC and its improved version the D-VBC (see Table \ref{VBCENERGY} for energy-comparison). The D-VBC state is therefore the best state for a doping level of around 10 percent.\\
For this state we measure the bond spin-exchange $S_i S_j$ the bond kinetic energy $c^\dagger_i c_j$ and the local on-site density $n_i$ for all bonds/sites and visualize them. For better visualization we subtract a finite value form all bonds/sites in order to see the patterns better. The results can be seen
in Fig. \ref{SPINKINDENS}. We observe the following: stronger spin-exchange takes place exactly at that pattern that was the input in the wave-function for the bond-modulation of the H-VBC. For the expectation value $S_{ij}=\langle S_i S_j \rangle$ we measure three ranges of values corresponding to $S_{ij}\approx -0.30$ for the bonds with the stronger spin-spin exchange, $-0.15 < S_{ij} < -0.11$ for the medium spin-spin exchange and $S_{ij}\approx -0.05$ for the bonds with the smallest spin-spin exchange. The kinetic energy 
follows a slightly different pattern, which is the combination of the one of the H-VBC and the inner hexagon. This resulting pattern was the original motivation to introduce a separate parametrization of this inner hexagon, it appears as well if $\chi_2=1$. The energy-gain is caused by this additional parameter is small (see Table \ref{VBCENERGY}. For the expectation value $T_{ij}=\langle c^{\dagger}_i c_j \rangle$ we measure $-0.15<T_{ij}<-0.14$ for the bonds with stronger kinetic energy and $T_{ij}\approx-0.09$ for those with less hopping. The densities measured are lowest in the hexagons, where $n_i=0.90$ and $n_i=0.92$ elsewhere.\\
Using earlier results on static holes and analyzing the resulting patterns we propose the dominant mechanism for forming this type of VBC at this doping level: 
(i) In the study of Dommange et al \cite{DOM} it was found that a static single impurity (empty site) induce a patten of bond spin-spin correlations, therefore inducing dimerization. Being the static bond on the corner-sharing triangle (on the kagome all sites are corner-sharing triangles) this strong-bond is at the bond facing the impurity.
(ii) In the same study the force between impurities was found to be repulsive, therefore holes are not expected to cluster.
The $t-J$ model has to balance its kinetic energy gain, with the spin-spin exchange. As the uniform or other flux states without VBCs having higher energy we know that the holes are most likely not evenly distributed. If there are certain hole trajectories, with a higher probability, then they may induce static density patterns, as we observe. In our particular case (of a little less than 10 percent doping) it means each ``Star-of-David" gets a little over one hole (1.2 holes to be precise). Assume now exactly one hole (being 1/12 doping) to loop along an inner hexagon: Because of (ii) this would induce the spin-spin exchange exactly in a pattern making up the H-VBC (see Fig. \ref{DIM}). If all ``Stars of Davids" have one hole, two holes will be closer than three lattice sites and never be further away from another hole in another hexagon than seven lattice sites. One realizes that the holes can only keep their optimal distance by staying in the same hexagon (see Fig. \ref{DIM}). When two holes have this closet distance, then in fact a move out of the inner hexagon of the star would decrease the distance by one lattice site, while 
staying on the hexagon leaves the distance constant in that move. Assume for now two holes in one hexagon: Now the two holes in the hexagon have a shortest distance of three lattice sites within the ``Star of David", therefore a confinement 
within the hexagon is no more favorable and eventually holes hop to other neighboring sites, and by the dimerization of this different part, violate the H-VBC patten. Therefore this particular VBC patten is not favorable anymore at  a higher doping level. This is consistent with our result in our earlier study. From the above we deduce an optimal doping level of 1/12 for this particular state. We have tested this hypothesis by VMC around this doping. If one varies the doping only slightly, there is only a very slight shift in the var. parameters. To distinguish them one needs to obtain the VMC energies to a very high accuracy. We could obtain such a curve for the 192 site system. For bigger systems it is currently beyond our computational power. See Fig. \ref{opt} for the result: We see a maximum in $\chi_1$ at extremely close 1/12 (within error-bars at 1/12) and a minimum in $\chi_2$ at the same point. The minimum and the maximum are slightly shifted. Note the relevance of the ratio of the two parameters, seting the scale between the weakest and strongest bonds, plotting this curve one sees the maximum at exactly 1/12 doping, with a ratio of around 1.3. 
Finally we discuss our result in the context of static impurities in actual compounds as e.g. herbertsmithite: When crystals are formed from a high-temperature phase in cooling down very slowly, it appears reasonable that in a forming background of a kagome lattice, there is some mobility of the impurities till the crystal freezes. If this scenario is indeed the case, the static impurity situation, would 
actually resemble a state with holes in hexagons when the doping level is at around $1/12$. The VBC as we see it in the dynamic case, would not exist in the same manner, as it relies on the fact
that all six inner hexagon sites have a higher impurity concentration, while once its frozen one site has $n_{imp}=1$ while all others have $n_{imp}=0$. With this picture in mind we compare with another investigation of
the Kagome lattice Heisenberg model with an static impurity based on series expansion by R. Singh \cite{SINGH}. In this work the frozen single impurity is assumed to be placed on various position within the star. Under the background of the freezing influenced by other impurities based on our observation, the frozen impurity resembles the situation in Fig. 2. a) in the mentioned work of R. Singh. The bulk state assumed to be grown in an very optimal way, should therefore consist of ``Star of David" units, as in Fig. 2 a) where the orientation of those building blocks randomly depends on where the impurity in the inner hexagon the impurity freezes. Therefore if our assumption that a slow freezing process of the material describes a $t-J$ model at an intermediate step is valid, we expect this to give it an additional symmetry compared to the result of R. Singh, in the form of six orientations of this particular building block (in Fig. 2. a) in \cite{SINGH}), being randomly distributed, while other cases (Fig. 2 b) and c) of the same reference) should not appear. The implications of such a state and its possible detection 
in an experiment deserves an careful investigation in its own right, and shall be the subject of future investigation.

\begin{table} 
\begin{tabular}{|l|l|l|l|l|} 
\hline 
State & Energy & $\chi_1$ & $\chi_2$ & $\chi_3$ \\ 
\hline 
Uniform&$-0.606(5)$ & - & - &  - \\
\hline 
Y-VBC&$-0.6083(7)$  & 1.00 & 0.80 & 1.10  \\ 
\hline
I-VBC&$-0.6076(3)$  & 1.04 & - &  - \\ 
\hline
%Z-VBC&$-0.606(9)$\\ 
%\hline
D-VBC&$-0.6099(2)$  & 1.19 & 0.94 &  -  \\ 
\hline
H-VBC&$-0.6098(8)$  & 1.22 & - &  -  \\
\hline 
\end{tabular}
\caption{Energies for various VBCs for a 192 site system at a doping of 9.4 percent (18 holes).} \label{VBCENERGY} 
\end{table}

To summarize, we have investigated the doped phase of the Kagome lattice $t-J$ model, testing several VBCs. Comparing our result for local bond and site measurements with the results on static holes in the Kagome lattice Heisenberg model we propose a  mechanism stabilizing VBC pattens in this model, and give an argument why the ``Star of David" is a good candidate around $1/12$ doping. We discussed the connection to static impurities and our analysis provides a consistent picture linking the Kagome lattice $t-J$ model and the diluted Heisenberg Kagome lattice model. Our work provides the missing link between the two models and suggests a route to a detailed understanding of actual compounds with impurities. We have given a detailed description of the correlations of our state, which are experimentally detectable.
Recently there have been several proposals of novel physics at the $1/6$ doping level in the Kagome lattice Hubbard model \cite{QIN,KI}. The connection or differences for this doping level for the Kagome lattice Hubbard model and Kagome lattice $t-J$ model (corresponding to each other at $U=4t^2/J$ and $V=0$) might be a very fruitful route for further exploration.\\

{\it Acknowledgments: } 
Supercomputer support was provided by the NIC, FZ J\"ulich under project
No. HHB00, through the HPC facilities of the RWTH-Aachen via the RV-NRW and by ITMC of TU-Dortmund.

\end{document}